\documentclass[showpacs,preprintnumbers,amsmath,amssymb]{revtex4}

\usepackage{graphicx}

\begin{document}

\title{Axially Symmetric Black Hole Skyrmions}

\author{Nobuyuki Sawado}
\email{sawado@ph.noda.tus.ac.jp}
\author{Noriko Shiiki}
\email{norikoshiiki@mail.goo.ne.jp}

\affiliation{Department of Physics, Tokyo University of Science, Noda, Chiba 278-8510, Japan}

\date{\today}

\begin{abstract}

It has been known that a $B=2$ skyrmion is axially symmetric. 
We consider the Skyrme model coupled to gravity and obtain static 
axially symmetric black hole solutions numerically. 
The black hole skyrmion no longer has integer baryonic charge but has fractional 
charge outside the horizon as in the spherically symmetric case. 
Therefore, the solution represents a black hole partially swallowing a deuteron.  
Recent studies of theories with large extra dimensions suggest an effective Planck scale 
of order a TeV and thus the deuteron black hole may be produced in the Linear Hadron 
Collider (LHC) in future. 

\end{abstract}

\pacs{04.70.Bw, 12.39.Dc, 21.60.-n \\
{\it Keywords}: Black holes, Skyrmions, Deuterons}

\maketitle


\section{Introduction}
It has been shown that the no-hair conjecture 
for black holes \cite{ruffini} is violated when 
some non-linear matter fields are considered. 
The first counter example was provided 
by Luckock and Moss \cite{luckock_moss} who found 
the Schwarzschild black hole with Skyrme hair. 
The presence of the horizon in the core of skyrmion 
unwinds the skyrmion, leaving fractional baryon charge 
outside the horizon. 
The full Einstein-Skyrme system was solved later by Droz 
{\it et al.} to obtain spherically symmetric black holes 
with Skyrme hair \cite{droz}. Other counter examples include 
static spherically symmetric black holes 
in the Einstein-Yang-Mills (EYM) \cite{volkov}, 
the Einstein-Yang-Mills-Dilaton (EYMD) \cite{torii,maison} 
and the Einstein-Yang-Mills-Higgs (EYMH) theory \cite{nair}. 
More interestingly, it has been also shown that these Einstein-Yang-Mills 
theories have static axially symmetric black hole solutions 
\cite{kleihaus,hartmann}. 

Motivated by the axially symmetric hairy black holes in 
Refs.~\cite{kleihaus,hartmann}, we shall study the Einstein-Skyrme 
model with axial symmetry. 
It has been shown that a $B=2$ skyrmion is axially symmetric and represents 
a deuteron~\cite{braaten}. 
Our model, therefore, provides a convenient framework to 
study the interactions between a deuteron and a black hole.  
By examining the baryon number of the solution, 
the absorption of the deuteron by the black hole is observed 
as in the spherically symmetric case.  
We expect our solutions are stable as skyrmions are topologically stable objects. 
 
Recent studies of theories with large extra dimensions indicate that a true Planck scale is 
of order a TeV and the production rate of black holes 
massive than the Planck scale become quite large~\cite{hamed,cavagria,banks}. 
Therefore, it may be possible to produce the deuteron black holes in the LHC 
in future. 

\section{The Model} 
The Skyrme model is an effective theory of QCD based on  
pion fields alone \cite{skyrme}. At low energy, the symmetry of 
the strong interaction is broken spontaneously and hence the 
Skyrme Lagrangian retains the chiral symmetry. 
The Skyrme model coupled to gravity is defined by     
\begin{eqnarray}
	{\cal L}={\cal L}_S + {\cal L}_G, \label{lg}
\end{eqnarray}
where
\begin{eqnarray*}
	{\cal L}_S &=& \frac{f_{\pi}^{2}}{16}g^{\mu\nu}{\rm tr}\left(U^{-1}
	\partial_{\mu}UU^{-1}\partial_{\nu}U\right) 
	+ \frac{1}{32a^{2}}g^{\mu\rho}g^{\nu\sigma}{\rm tr}\left([U^{-1}\partial_{\mu}
	U, U^{-1}\partial_{\nu}U][U^{-1}\partial_{\rho}U, U^{-1}
	\partial_{\sigma}U]\right), \label{ls} \\
	{\cal L}_G &=& \frac{1}{16\pi G}R . \label{}
\end{eqnarray*} 
Let us introduce an ansatz for the metric given in Ref.~\cite{kleihaus}
\begin{eqnarray}
	ds^{2} = -fdt^{2} + \frac{m}{f}(dr^{2} + r^{2} d\theta^{2}) + \frac{l}{f}
	r^{2} \sin^{2}\theta d\varphi^{2},
\end{eqnarray}
where $f=f(r,\theta), \,\, m=m(r,\theta), \,\,{\rm and}\,\, l=l(r,\theta)$. \\

The axially symmetric Skyrme field can be parameterized by  
\begin{eqnarray}
	U=\cos F(r,\theta)+i{\vec \tau}\cdot{\vec n}_R\sin F(r,\theta),
	\label{chiral}
\end{eqnarray}
with ${\vec n}_R = (\sin\Theta\cos n\varphi, \sin\Theta\sin n\varphi, \cos\Theta)$. 
In terms of $F$ and $\Theta$, the Lagrangian (\ref{lg}) has the form 
\begin{eqnarray}
	{\cal L}_S={\cal L}_{S}^{(1)}+{\cal L}_{S}^{(2)}, \label{}
\end{eqnarray}
where  
\begin{eqnarray}
	{\cal L}_S^{(1)}& =& -\frac{f_{\pi}^{2}f}{8m}\left\{(\partial_{r}F)^{2}
	+\frac{1}{r^{2}}(\partial_{\theta}F)^{2}+\left[(\partial_{r}\Theta)^{2}+\frac{1}{r^{2}}
	(\partial_{\theta}\Theta)^{2}\right]\sin^{2}F+\frac{n^{2}}{r^{2}\sin^{2} \theta}
	\frac{m}{l}\sin^{2}{\Theta}\sin^{2}F \right\}, \nonumber \\
	{\cal L}_S^{(2)}&=&-\frac{1}{2a^{2}r^{2}}\left(\frac{f}{m}\right)^{2}
	\left\{(\partial_{r}F\partial_{\theta}\Theta -\partial_{\theta}F
	\partial_{r}\Theta )^{2}+\frac{n^{2}}{\sin^{2}\theta }\frac{m}{l}
	\left[(\partial_{r}F)^{2}+\frac{1}{r^{2}}(\partial_{\theta}F)^{2}\right]
	\sin^{2}\Theta \right\}\sin^{2}F   \nonumber \\ 
	&& -\frac{1}{2a^{2}r^{2}}\left(\frac{f}{m}\right)^{2}\left\{ 
	\frac{n^{2}}{\sin^{2}\theta}\frac{m}{l}\left[(\partial_{r}\Theta )^{2}
	+\frac{1}{r^{2}}(\partial_{\theta}\Theta)^{2}\right]\sin^{2}F\sin^{2}\Theta  
	\right\}\sin^{2}F . \nonumber
\end{eqnarray}
Since we are interested in $B=2$, we shall take the winding number $n=2$. 

The baryon current in curved spacetime is obtained by 
taking the spacetime covariant derivative $\nabla_{\mu}$, 
\begin{eqnarray}
	b^{\mu}=\frac{1}{24\pi^{2}}\epsilon^{\mu\nu\rho\sigma}
	{\rm tr}(U^{-1}\nabla_{\nu}UU^{-1}\nabla_{\rho}U
	U^{-1}\nabla_{\sigma}U) .\label{}
\end{eqnarray}
The baryon number then is given by integrating $b^{0}$ 
over the hypersurface $t=0$, 
\begin{eqnarray*}
	B&=&\int drd\theta d\varphi \sqrt{g^{(3)}}\; b^{0}  \\
	&=& -\frac{1}{\pi}\int dr d\theta\; (\partial_{r}F
	\partial_{\theta}\Theta-\partial_{\theta}F\partial_{r}\Theta)
	\sin\Theta(1-\cos 2F) \\ 
	&=& -\frac{1}{\pi}\int dF \wedge d\Theta \sin \Theta (1-\cos 2F) \\ 
	&=& \left[ \frac{1}{2\pi}(2F-\sin 2F)\cos \Theta \right]_{F_{0},
	\Theta_{0}}^{F_{1},\Theta_{1}},
\end{eqnarray*}
where $(F_{0},\Theta_{0})$ and $(F_{1},\Theta_{1})$ are the values at the inner 
and outer boundary, respectively. 
In flat spacetime~\cite{braaten}, we have
\begin{eqnarray}
	(F_{0},\Theta_{0})=(\pi,0)\;\; {\rm and} \;\;(F_{1},\Theta_{1})=(0,\pi) , \label{}
\end{eqnarray}
which gives $B=2$. In the presence of a black hole, the integration 
should be performed from the horizon to infinity, which changes 
the values of $F_{0}$ and allows the $B$ to be fractional. 

The horizon mass can be derived from the first law of 
a black hole in the isolated horizon framework~\cite{ashtekar}. 
For this purpose, let us introduce the area element 
\begin{eqnarray}
	 dA &=& d\theta d\varphi \sqrt{g_{\theta\theta}
	g_{\varphi\varphi}} \nonumber \\
	&=&2\pi r^{2} \sin\theta d\theta
	\frac{\sqrt{lm}}{f},  \label{}
\end{eqnarray}
and thus 
\begin{eqnarray}
	A(r)=2\pi r^{2}\int_{0}^{\pi} \sin\theta d\theta \frac{\sqrt{lm}}{f}. \label{}
\end{eqnarray}
An effective horizon radius is then defined by 
\begin{eqnarray}
	\rho_{h}=\sqrt{\frac{A(r_{h})}{4\pi}} \label{}
\end{eqnarray}
where $r=r_{h}$ is a horizon of the black hole.
The surface gravity is given by \cite{e_weinberg}, 
\begin{eqnarray}
	\kappa^{2} &=&-\frac{1}{4}g^{00}g^{ij}(\partial_{i}g_{00})
	(\partial_{j}g_{00}) \nonumber \\
	&=& \frac{1}{4m}\left\{(\partial_{r}f)^{2}
	+\frac{1}{r^{2}}(\partial_{\theta}f)^{2}\right\}.  \label{kappa}
\end{eqnarray}
According to the first law of a static black hole, one obtains the 
horizon mass 
\begin{eqnarray*}
	M_{h}&=&\int_{0}^{\rho_{h}}\frac{\kappa}{8\pi}\frac{dA(\rho_{h}')}
	{d\rho_{h}'}d\rho_{h}' \nonumber \\
	&=& \int_{0}^{\rho_{h}}\kappa \rho_{h}' d\rho_{h}',
\end{eqnarray*}
which can be evaluated numerically. 

\section{Boundary Conditions}
Let us consider the boundary conditions for the chiral fields and metric functions 
with help of Ref.~\cite{kleihaus}. 
At the horizon $r=r_{h}$, the time-time component of the metric satisfies 
\begin{eqnarray}
	g_{tt}=-f(r_{h},\theta)=0 . \label{}
\end{eqnarray}
Regularity of the metric at the horizon requires  
\begin{eqnarray}
	m(r_{h},\theta)=l(r_{h},\theta)=0 . \label{}
\end{eqnarray}
The boundary conditions for $F(r,\theta)$ and $\Theta(r,\theta)$ at the horizon 
are obtained by expanding them at the horizon and inserting into the field equations 
derived from $\delta {\cal L}_{S}/\delta F=0$ and $\delta {\cal L}_{S}/\delta \Theta=0$ 
respectively, 
\begin{eqnarray}
	\partial_{r}F(r_{h},\theta)=\partial_{r}\Theta(r_{h},\theta)=0.\label{}
\end{eqnarray} 
The condition that the spacetime is asymptotically flat requires
\begin{eqnarray}
	 f(\infty,\theta)=m(\infty,\theta)=l(\infty,\theta)=1 . \label{}
\end{eqnarray}
The boundary conditions for $F$ and $\Theta $ at infinity remain the same 
as in flat spacetime      
\begin{eqnarray}
	F(\infty,\theta)=0,\; \partial_{r}\Theta(\infty,\theta)=0 . \label{}
\end{eqnarray}
For the solution to be axially symmetric, we have 
\begin{eqnarray}
 	&&\partial_{\theta}f(r,0)
	=\partial_{\theta}m(r,0)
	=\partial_{\theta}l(r,0)=0 , \\
	&&\partial_{\theta}f\left(r,\frac{\pi}{2}\right)
	=\partial_{\theta}m\left(r,\frac{\pi}{2}\right)
	=\partial_{\theta}l\left(r,\frac{\pi}{2}\right)
	=0 . \label{}
\end{eqnarray}
Likewise for $F$, 
\begin{eqnarray}
	\partial_{\theta}F(r,0)=\partial_{\theta}F\left(r,\frac{\pi}{2}\right)=0 . \label{}
\end{eqnarray}
Regularity on the axis and axisymmetry impose the boundary conditions on $\Theta$ as  
\begin{eqnarray}
	\Theta(r,0)=0, \; \Theta\left(r,\frac{\pi}{2}\right)=\frac{\pi}{2} . \label{}
\end{eqnarray}

\section{Numerical Results and Discussions}
\begin{figure}
\includegraphics[height=9.5cm, width=12.5cm]{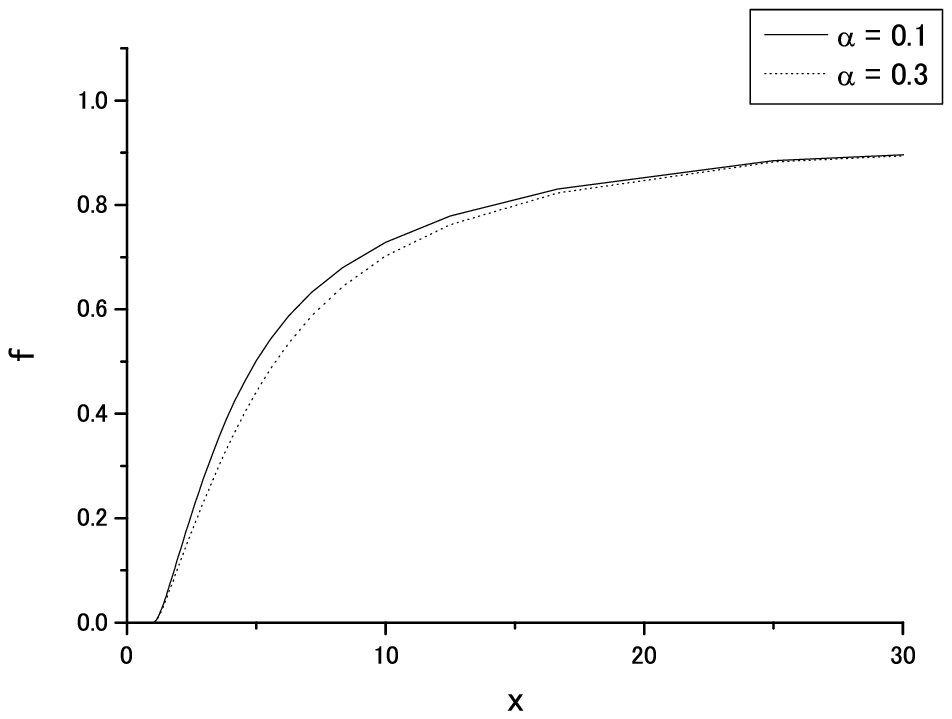}
\caption{\label{fig:rf} The metric function $f$ as a function of $x$ with 
$\theta = \pi/4$ and $x_{h}=1.0$. } 
\includegraphics[height=9.5cm, width=12.5cm]{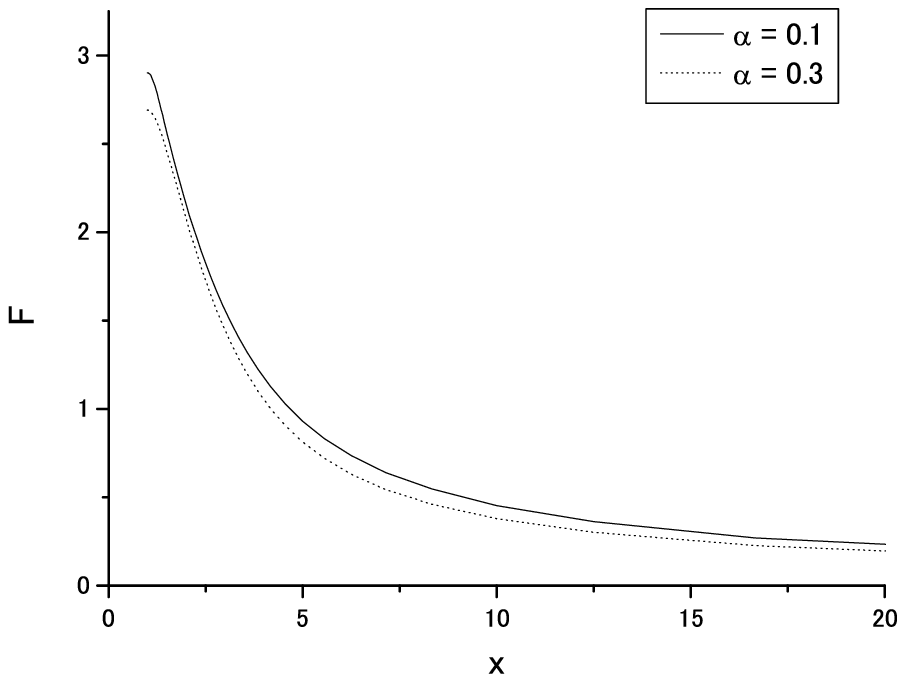}
\caption{\label{fig:rfs} The Skyrme function $F$ as a function of $x$ 
with $\theta = \pi/4$ and $x_{h}=1.0$.}
\end{figure}

\begin{figure}
\includegraphics[height=9.5cm, width=12.5cm]{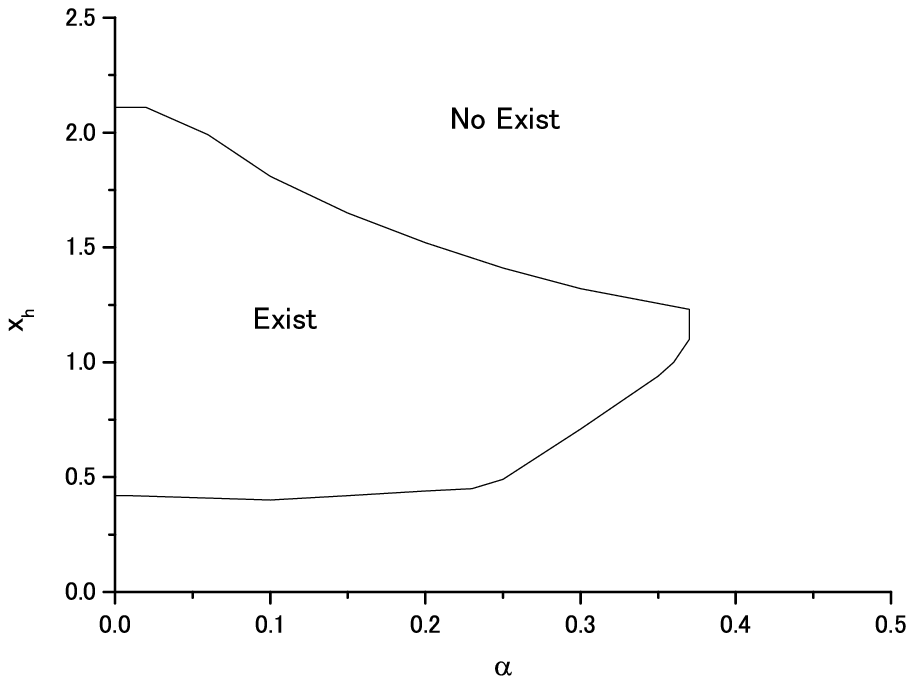}
\caption{\label{fig:krh} The domain of existence of the 
solution. For $\alpha > 0.37$, there exists no non-trivial 
solution.}
\includegraphics[height=9.5cm, width=12.5cm]{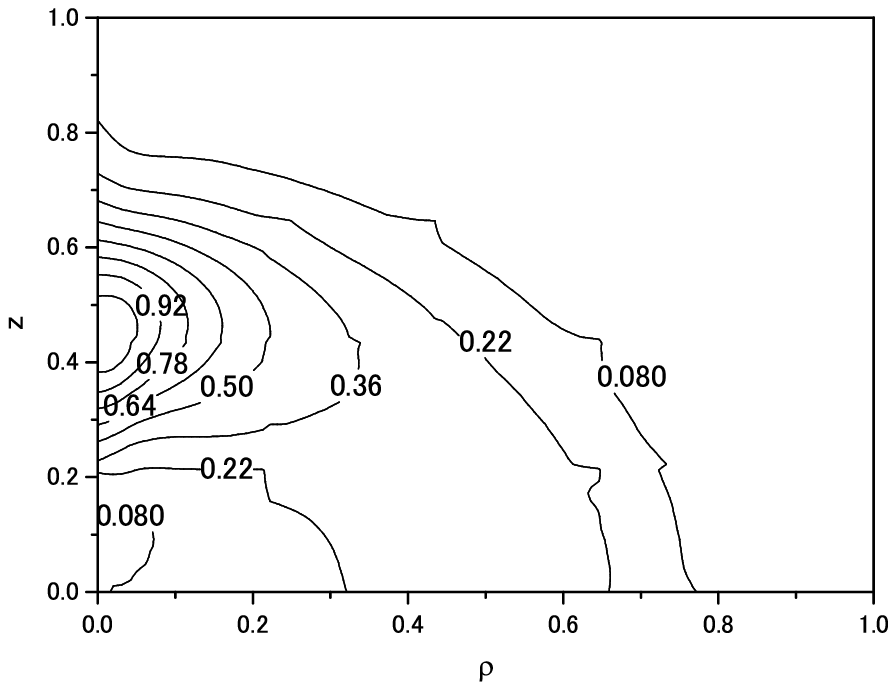}
\caption{\label{fig:red} A contour plot of the energy density $\epsilon$ 
in cylindrical coordinates $\rho$ and $z$ with $x_{h}=0.8$. } 
\end{figure}

\begin{figure}
\includegraphics[height=9.5cm, width=12.5cm]{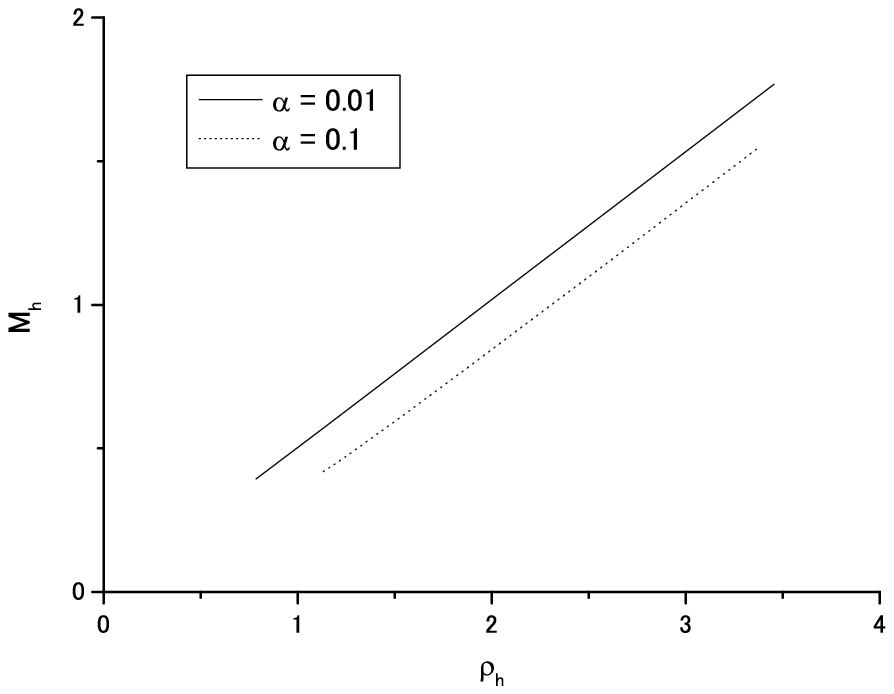}
\caption{\label{fig:rhmh} The horizon mass $M_{h}$ as a function of $\rho_{h}$. } 
\includegraphics[height=9.5cm, width=12.5cm]{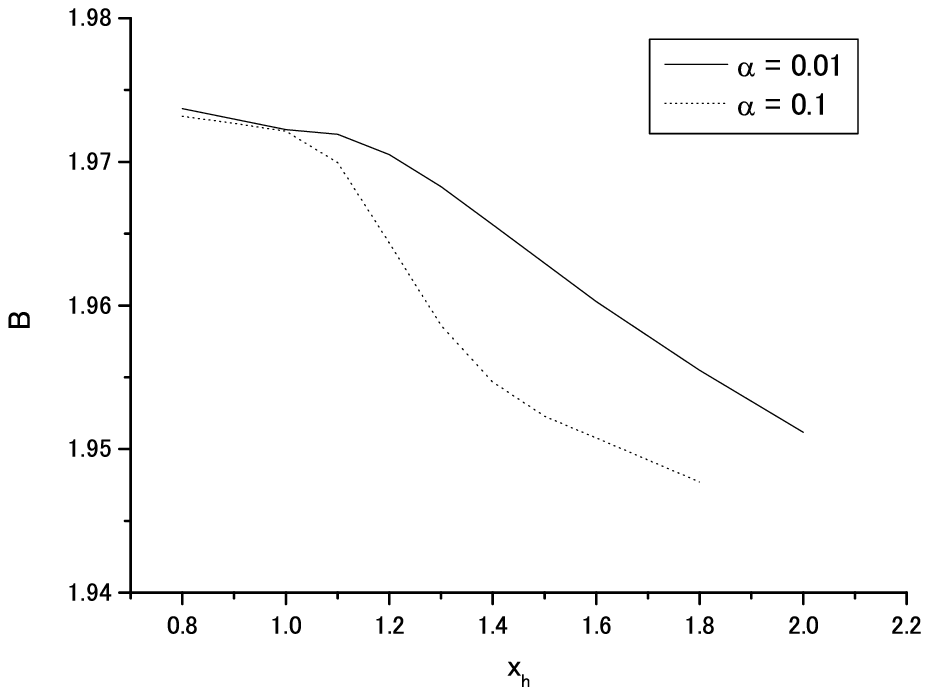}
\caption{\label{fig:rhb} The dependence of the baryon number 
on the size of the horizon.}
\end{figure}

For the purpose of numerical computation, we shall introduce 
a dimensionless radial coordinate $x$ and coupling constant $\alpha$,  
\begin{eqnarray}
	x=af_{\pi}r , \;\;  \alpha = \pi G f_{\pi}^{2} .\label{}
\end{eqnarray}
Then, in this system, free parameters are only $x_{h}$ and $\alpha$. 
Fig.~\ref{fig:rf} shows metric function $f$ for $\alpha=0.1, 0.3$ 
with $\theta = \pi/4$. For $\alpha < 0.1$,   
the results are similar to that of $\alpha = 0.1$. Other metric functions $l$ and $m$ 
exhibit similar shapes as $f$. The $\theta$ dependence 
of the metric functions are very small.   
The Skyrme function $F$ is shown in Fig.~\ref{fig:rfs}. 
The domain of existence of the solutions in the parameter space is shown in Fig.~\ref{fig:krh}. 
For $\alpha \gtrsim 0.37$, there exists no solution since the chiral  
fields become too massive for the black hole to support outside the horizon. 
It is also observed that the black hole has a finite minimum size unlike the 
spherically symmetric case. Hence one can not recover regular solutions as the limit of 
zero horizon size.   
We show the energy density of the Skyrme fields $(\epsilon = -T_{0}^{0})$
in Fig.~\ref{fig:red}. The density is dumbbell in shape with the highest along $z$-axis 
while in flat spacetime it is toroidal. It is interesting to see how the gravitational 
interaction affects the shape of the skyrmion. 
The dependence of the horizon mass $M_{h}$ on $\rho_{h}$ is shown in Fig.~\ref{fig:rhmh}. 
As $\alpha $ becomes smaller, the black hole approaches to the Schwarzschild black hole with 
$M_{S}=\rho_{h}/2$. However, according to the argument in Ref.~\cite{ashtekar}, 
it does not exceed $M_{S}$ for all values of $\rho_{h}$.  
This figure indicates that our solutions are stable since they lie on a high-entropy 
branch with maximal entropy~\cite{maeda}. From the study of the spherically symmetric 
case in Ref.~\cite{luckock_moss}, we suspect that there also exist unstable low-entropy 
branches although we could not find them by the over-relaxation 
numerical scheme. 
Fig.~\ref{fig:rhb} shows the dependence of the baryon number on $\alpha$ and $x_{h}$. 
It is observed that the baryon number gets more absorbed by the black hole 
in increase of the size of the black hole and the coupling constant. 

Finally, theories with extra dimensions bring us an interesting possibility that the deuteron 
black holes could be produced in the LHC by collision of two protons. 
The produced black holes then should be rotating and therefore it will be worth 
studying rotating deuteron black holes.     
The inclusion of gauge fields should also be necessary to study electrically charged 
deuteron black holes~\cite{moss_shiiki}.  

\begin{center}
	{\bf Acknowledgements}
\end{center}
We are grateful to K. Maeda and T. Torii for many useful discussions. 

\begin{thebibliography}{99}

\bibitem{ruffini}
R. Ruffini and J. A. Wheeler, Physics Today {\bf 24} (1971) 30.  

\bibitem{luckock_moss}
H. Luckock and I. G. Moss, Phys. Lett. {\bf B176} (1986) 341;  
H. Luckock, {\it String theory, quantum cosmology etc.} Eds. 
H. J. de Vega and N. Sanchez (world scientific, 1987). 

\bibitem{droz}
S. Droz, M. Heusler and N. Straumann, Phys. Lett. {\bf B268} (1991) 371. 

\bibitem{volkov}
M. S. Volkov and D. V. Galtsov, JETP Lett. {\bf 50} (1989) 346. 

\bibitem{torii}
T.Torii and K. Maeda, Phys. Rev. D {\bf 48} (1993) 1643;

\bibitem{maison}
G. V. Lavrelashvili and D. Maison, Nucl. Phys. {\bf B410} (1993) 407. 

\bibitem{nair}
K-M. Lee, V. P. Nair and E. J. Weinberg, Phys. Rev. {\bf D45} (1992) 2751. 

\bibitem{kleihaus}
B. Kleihaus and J. Kunz, Phys. Rev. Lett. {\bf 79} (1997) 1595.

\bibitem{hartmann} 
B. Hartmann, B. Kleihaus and J. Kunz, Phys. Rev. {\bf D65} (2002) 024027.

\bibitem{braaten}
E. Braaten and L. Carson, Phys. Rev. {\bf D38} (1988) 3525.

\bibitem{hamed}
N. Arkani-Hamed, S. Dimopoulos and G. Dvali, Phys. Lett. {\bf B429} (1988) 263

\bibitem{cavagria}
M. Cavaglia, Int. J. Mod. Phys. {\bf A18} (2003) 1843

\bibitem{banks}
T. Banks and W. Fischler, {\it hep-th}/9906038

\bibitem{skyrme}
T. H. R. Skyrme, Proc. Roy. Soc. Lond. {\bf A260} (1961) 127.

\bibitem{ashtekar}
A. Ashtekar, A. Corichi and D. Sudarsky, Class. Quant. Grav. {\bf 18} (2001) 919.

\bibitem{e_weinberg}
S. A. Ridgway and E. J. Weinberg, Phys. Rev. {\bf D52} (1995) 3440.


\bibitem{maeda}
K. Maeda, T. Tachizawa, T. Torii and T. Maki, Phys. Rev. Lett. {\bf 72} 
(1994) 450; T. Torii, K. Maeda and T. Tachizawa, Phys. Rev. {\bf D51} 
(1995) 1510.



\bibitem{moss_shiiki}
I. G. Moss, N. Shiiki and E. Winstanley, Class. Quant. Grav. {\bf 17} 
(2000) 4161

\end{thebibliography}
\end{document}